\documentclass[twocolumn,showpacs,preprintnumbers,amsmath,amssymb]{revtex4}

\usepackage{graphicx}
\usepackage{dcolumn}
\usepackage{bm}

\begin{document}
\title{Thermoelectricity of URu$_{2}$Si$_{2}$: giant Nernst effect in the hidden-order state}
\author{Romain Bel$^{1}$, Hao Jin$^{1}$, Kamran Behnia$^{1}$, Jacques Flouquet$^{2}$ and Pascal
Lejay$^{3}$}
 \affiliation{(1)Laboratoire de Physique Quantique(CNRS), ESPCI, 10 Rue de Vauquelin,
75231 Paris, France \\ (2)DRFMC/SPSMS,  Commissariat \`a l'Energie
Atomique, F-38042 Grenoble, France\\
(3)Centre de Recherche sur les Tr\`es Basses Temp\'eratures(CNRS),
F-38042 Grenoble, France}

\date{July 6, 2004}

\begin{abstract}
We present a study of Nernst and Seebeck coefficients in the
heavy-fermion compound URu$_{2}$Si$_{2}$, which hosts a phase
transition of unsettled origin. A giant Nernst signal of
unprecedented magnitude was found to emerge in the ordered state.
Moreover, our analysis of the Seebeck and Hall data indicates that
the ordering leads to a sudden increase in the entropy per
itinerant electron and to a drastic decrease in the scattering
rate.
\end{abstract}


\maketitle

\begin{figure}
{\includegraphics{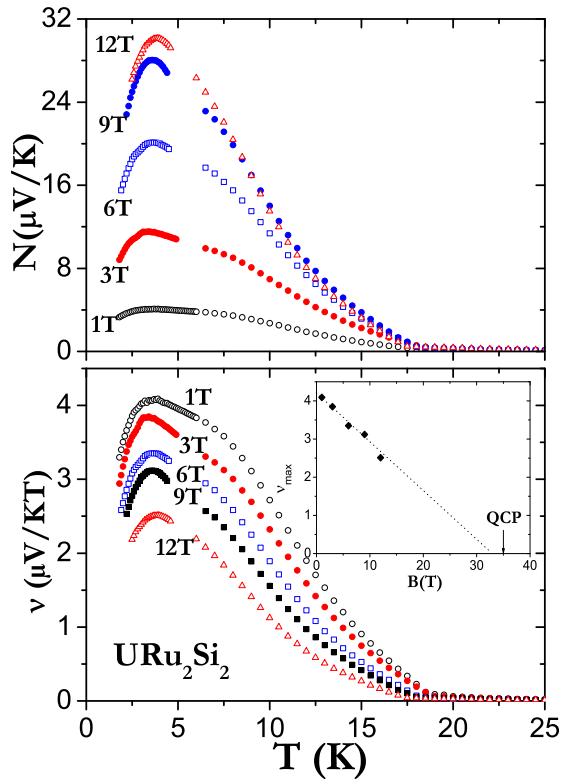}} \caption{\label{fig1}Temperature
dependence of the Nernst signal (upper panel) and coefficient
(lower panel) for different magnetic fields in URu$_{2}$Si$_{2}$.
Upper inset shows the temperature dependence of the transverse to
longitudinal voltages produced by a longitudinal thermal gradient.
Lower inset presents the field dependence of the peak value of
$\nu$. The arrow presents the field corresponding to the
destruction of the hidden order\cite{kim}.}
\end{figure}

Among heavy-fermion superconductors, URu$_{2}$Si$_{2}$ is
distinguished by the presence of a mysterious electronic order
below T$_{0}$=17.5K\cite{palstra,schlabitz,maple}. A large amount
of magnetic entropy ($S_{mag} \sim 0.2Rln2$) is lost in this phase
transition\cite{palstra}. Nevertheless, the intensity of the
emerging magnetic Bragg peaks imply an anti-ferromagnetic order
with a very weak magnetic moment($\sim 0.03
\mu_{B}$/U)\cite{broholm}. With the application of pressure, the
the magnetic Bragg peaks intensify and a conventional AF ground
state emeges\cite{amitsuka,bourdarot1}. At ambient pressure, the
small moment resolved cannot account by itself for the amount of
entropy lost at the transition. However, ordering opens a sizeable
gap in both spin and charge excitations. This unusual case of
large macroscopic anomalies leading to a tiny magnetic moment has
nourished an extensive investigation during the last two decades.
Many proposals\cite{barzykin,santini,sikkema,okuno,ikeda,chandra1}
regarding the nature of what is now commonly called the hidden
order of URu$_{2}$Si$_{2}$\cite{shah} have emerged. Some of these
invoked exotic order parameters such as
three-spin-correlator\cite{barzykin}, quadrupolar ordering of
localized moments\cite{santini} or an unconventional
spin-density-wave\cite{ikeda}. In some others, weak
antiferromagnetism is explained by considering the dual (i.e.
localized and itinerant) character of the $5f$ electron in the
context of a Fermi surface with nesting\cite{sikkema,okuno,fomin}.

Several recent experiments indicate that the hidden order and
Large-Moment-Anti-ferromagnetism(LMAF) are in
competition\cite{yokoyama,motoyama,matsuda}. Anomalies indicating
a phase transition between two thermodynamically-distinct states
have been reported in thermal expansion studies\cite{motoyama}.
According to $^{29}$Si NMR measurements\cite{matsuda}, the
hidden-order state is spatially inhomogeneous with coexisting LMAF
and paramagnetic regions. The apparent weakness of
anti-ferromagnetism would be a consequence of the tiny fraction of
the volume occupied by the magnetically-ordered electrons at
ambient pressure. This result provides a strong case for an
electronic phase separation between the two order
parameters\cite{chandra2}.

In this paper, we report on the thermoelectricity of
URu$_{2}$Si$_{2}$ which provides new pieces to this puzzle. The
hidden-order state was found to host a Nernst coefficient of an
exceptionally large magnitude. This giant Nernst coefficient
decreases smoothly with the application of a magnetic field and
the extrapolation of data indicates that it would not survive the
destruction of the hidden order at high magnetic fields\cite{kim}.
On the other hand, the onset of the ordering is accompanied with
an enhancement of the absolute value of the Seebeck coefficient.
This is in sharp contrast with the concomitant decrease in the
electronic specific heat\cite{palstra,maple,dijk}. We will argue
that it can be understood by taking into account the change in the
carrier density induced by the transition.

Single crystals of URu$_{2}$Si$_{2}$ were prepared in a three arc
furnace under purified argon atmosphere and annealed under UHV for
one week at 1050$^{\circ}$C. Four longitudinal and two lateral
contacts were made on each crystal. Nernst and Seebeck
coefficients were measured using a one-heater-two-thermometers
set-up which allowed us to measure thermal and electric
conductivities as well as the Hall coefficient of the sample in
the same conditions. The results reported here were reproduced on
two different samples.

Fig.1 presents the temperature dependence of the Nernst effect for
various magnetic fields in URu$_{2}$Si$_{2}$. The upper panel
displays the ratio of the transverse electric field to the
longitudinal thermal gradient (N=-E$_{y}/\nabla_{x}$T), while the
lower panel presents the Nernst coefficient($\nu$=N/B). As seen in
the figure, below T$_{0}\sim$ 17.5K, the Nernst signal (which is
negligibly small above this temperature) begins to grow steadily
and attains its maximum value at T$\sim$3K. This Nernst signal
presents an sublinear field-dependence as indicated by the
field-induced decrease of $\nu$ seen in the lower panel of the
figure. Plotting $\nu_{max}$ as a function of magnetic field, one
can estimate the field scale associated with the destruction of
the giant Nernst signal (see the lower inset of the figure). A
simple linear extrapolation of $\nu_{max}(B)$ to higher fields
puts this field at B$\sim$ 32T. Since this is comparable to the
magnitude of the field required to destroy the hidden order($\sim$
35T), it is safe to assume that the giant Nernst signal is another
property of the ordered state.

The magnitude of the Nernst coefficient is remarkably large. In
the zero-field limit, it attains $\nu_{max}\sim $4.2 $\mu$V/KT,
and easily exceeds what is reported in any other metal\cite{wu}
including NbSe$_{2}$\cite{bel1}($\nu_{max}\sim$ -0.12$\mu$ V/KT)
or CeCoIn$_{5}$\cite{bel2}($\nu_{max}\sim$ -0.95$\mu$ V/KT). At
B=12T, the signal is even larger than what is observed in the
vortex state of any superconductor.  In the zero-field limit,
however, the Nernst coefficient in a cuprate superconductor,
Ba$_{2}$Sr$_{1.5}$La$_{0.5}$CuO$_{6}$ is reported to attain a
comparable magnitude ($\nu_{max}\sim$ 4.5 $\mu$V/KT)\cite{wang}.

\begin{figure}
\resizebox{!}{0.4\textwidth}{\includegraphics{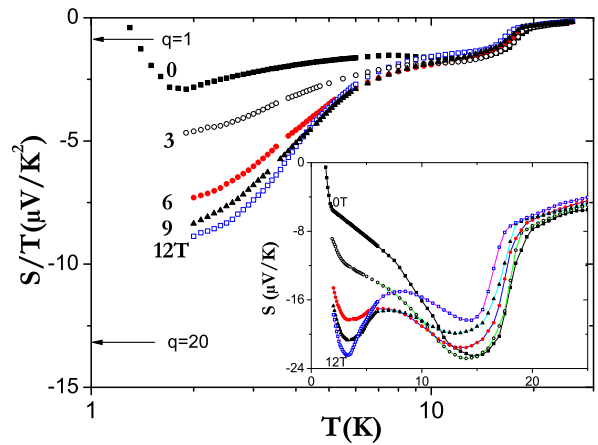}}
\caption{\label{fig2} Temperature-dependence of the Seebeck
coefficient divided by temperature as a function of temperature
for different magnetic field. The arrows point to the expected
$S/T$ given the electronic specific heat of URu$_{2}$Si$_{2}$ for
different carrier densities (see text). The inset shows the same
data in a more traditional fashion.}
\end{figure}

The phase transition which leads to the emergence of a Nernst
signal of such a magnitude affects also the [longitudinal]
thermopower\cite{sakurai}. As seen in Fig.2, the Seebeck
coefficient, $S$, presents a sharp anomaly at T$_{0}$ and becomes
strongly field dependent in the hidden order state. As seen in the
inset of Fig.2, a new low-temperature maximum in $S(T)$ emerges in
fields exceeding 6T. However, in order to resolve the magnitude of
temperature-linear thermopower in the zero temperature limit, a
plot of $S/T$ as a function of temperature\cite{behnia} is more
instructive. As seen in the main panel of the figure, the
application of a magnetic field leads to a steady increase in the
magnitude of $S/T$ down to the lowest temperatures investigated.
In many metals, the absolute value of a dimensionless ratio,
$q=\frac{S}{T \gamma}N_{Av} e$, linking the Seebeck coefficient to
the electronic specific heat ($\gamma= C_{el}/T$) is close to
unity\cite{behnia} in the T=0 limit. [Here $N_{Av}$ is the
Avogadro number and $e$ is the elementary charge.] This is not the
case of URu$_{2}$Si$_{2}$. Taking the zero-field values of $S/T$
(-2.8$\mu$V/K$^{2}$) and $\gamma$ (60 mJ/molK$^{2}$)\cite{dijk})
at the onset of superconductivity yields $q= -4.5$. The
enhancement of the absolute value of $S/T$ with the application of
the magnetic field (which leaves $\gamma$ unchanged in this
range\cite{dijk}) leads to an even larger $q$ at higher
fields(q$\sim$-14 at B=12T). The low carrier density in
URu$_{2}$Si$_{2}$ provides a natural explanation for this large
magnitude of $q$. The conversion factor, $N_{Av}e$, between
thermopower and specific heat assumes that there is a single
carrier per formula unit. Trivially, a proportionally larger $q$
is expected when the density of carriers is lower than this. The
case for a small carrier density in the hidden-order state of
URu$_{2}$Si$_{2}$ is supported by several other experimental
observations. It can be directly deduced from the magnitude of the
Hall coefficient in the zero-temperature limit which was found to
be R$_{H}=9.5 \times 10^{-3} cm^{3}/C$ in agreement with previous
studies\cite{schoenes,dawson}. While at finite temperatures, the
magnitude and the temperature-dependence of the Hall coefficient
of the Heavy-Fermion compounds depend largely on skew
scattering\cite{fert}, the extraordinary Hall effect vanishes in
the zero-temperature limit. Assuming a one-band model, the
magnitude of the Hall coefficient yields a carrier density of 0.05
holes per Uranium atom, as previously
reported\cite{schoenes,dawson}. Such a low carrier density implies
a Fermi surface occupying a small portion of the Brillouin zone
which is consistent with the results of band calculations based on
the spin-polarized Dirac equation\cite{yamagami}. Direct evidence
for  small [and almost spherical] Fermi surfaces comes from de
Haas- van Alphen and Shubnikov-de Haas
measurements\cite{bergemann,ohkuni,keller}. In the most complete
set of these studies, Keller \emph{et al.}\cite{keller} identified
four different orbits with the largest occupying less than 5
percent of the Brillouin zone.
\begin{figure}
\resizebox{!}{0.4\textwidth}{\includegraphics{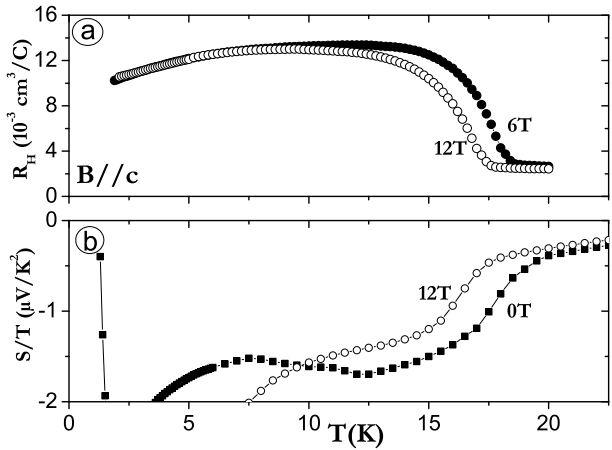}}
\caption{\label{fig3}a) Temperature-dependence of the Hall
coefficient. b) Same for thermopower divided by temperature. Note
the concomitant jumps in the vicinity of T$_{0}$.}
\end{figure}

Analysis of the behavior of thermopower and Hall coefficient in
the vicinity of T$_{0}$ leads to a subtle reconsideration of what
occurs to itinerant electrons at T$_{0}$. As seen in Fig.3 which
compares the anomalies in $R_{H}$ and in $S/T$, the absolute value
of both shows a step-like increase in a rather narrow temperature
window($\sim$ 3K). The three-fold \emph{increase }in the magnitude
of $S/T$ contrasts with the concomitant \emph{decrease} of
$\gamma$ which passes from 180 mJ/K$^{2}$mol above T$_{0}$ to 60
mJ/K$^{2}$mol below. Therefore,  the transition leads to a
reduction of the entropy per volume yet an enhancement of the
entropy associated to each carrier. This is confirmed by the sharp
change in the Hall coefficient at T$_{0}$. The five-fold increase
in $R_{H}$ corresponds to an even larger drop in carrier density
since a sizeable fraction of the Hall signal at this temperature
comes from skew scattering. Since the magnetic susceptibility is
barely affected by the transition\cite{schoenes,dawson}, the
contribution of skew scattering to Hall effect between 20K and 15
K remains unchanged. An analysis based on the Fert-Levy
model\cite{fert} yields a carrier density of 0.4 holes per U above
T$_{0}$ and ten times less below\cite{dawson}. The three-fold
decrease in $\gamma$ at T$_{0}$ appears, therefore, to mask two
opposing tendencies: a ten-fold reduction in carrier density and a
three-fold increase in entropy per carrier. In other words, while
90 percent of the carriers condense at the transition, those which
survive carry a larger entropy.

Thus, instead of being a light-weight heavy fermion, URu$_2$Si$_2$
in the hidden-order state emerges from this analysis as a system
with a very low density of carriers but a fairly large amount of
entropy per carrier. A carrier density of 0.05 holes per f.u.
implies $q\sim -20$. For $\gamma \sim$ 60mJ/molK$^{2}$, this
yields an expected $S/T= -12.5 \mu$V/K$^{2}$. As seen in Fig.2,
the low-temperature magnitude [of the absolute value] of $S/T$ at
zero-magnetic field is considerably lower than this. With the
application of magnetic field, $S/T$ is enhanced and approaches
the expected magnitude without attaining it. Since R$_H$ (and
consequently the carrier density) shows little variation with
magnetic field in this range, the strong change in $S/T$ with
magnetic field at finite temperature seems to reflect the field
dependence of the scattering time.

Let us now consider the emergence of a giant Nernst signal in such
a context. For this, it is useful to focus on a very simple
relationship which relates the Nernst coefficient to the Hall
angle in the Boltzmann picture and recently put forward by
Oganesyan and Usshishkin\cite{oganesyan}:
\begin{equation}\label{1}
N= \frac{\pi^{2}}{3}\frac{k_{B}^{2}T}{e}\frac{\partial
\Theta_{H}}{\partial \epsilon}|_{ \epsilon_{F}}
\end{equation}

Since the Hall angle,
$\Theta_{H}=\frac{\sigma_{xy}}{\sigma_{xx}}$, is a convenient
measure of the scattering time, $\tau$ (namely,
$\Theta_{H}=\omega_{c} \tau$, where $\omega_{c}$ is the cyclotron
frequency), then, Eq. 1 states that the Nernst signal measures the
energy-dependence of the scattering time at the Fermi energy. The
latter quantity is small in common metals, leading to a negligible
Nernst signal, a phenomenon also called Sondheimer
cancellation\cite{wang}. The situation becomes more complicated in
multi-band metals where the thermal gradient creates a counterflow
of carriers with opposite signs creating a finite Nernst signal.
Such an ambipolar Nernst effect was observed in NbSe$_{2}$ which
happens to present a sign-changing Hall coefficient in its Charge
Density Wave state\cite{bel1}. The giant Nernst signal in
URu$_{2}$Si$_{2}$, however, is more than one order of magnitude
larger. Moreover, it emerges in presence of a large and
\emph{non-vanishing} Hall angle. In order to explore other
possible sources of enhanced Nernst signal in a Boltzmann picture,
let us make a crude approximation and replace the energy
derivative $\frac{\partial \Theta_{H}}{\partial \epsilon}$ at the
Fermi level by $\frac{ \Theta_{H}}{\epsilon_F}$. In this case,
Eq.1 yields:

\begin{equation}\label{2}
N \approx 283 \frac{\mu V }{K} \times \Theta_H \times
\frac{k_{B}T}{\epsilon_F}
\end{equation}

A number of elementary insights are provided by this simple
expression. First of all, compared to simple metals, the
heavy-fermion compounds are expected to display an enhanced Nernst
signal simply because of the small magnitude of their Fermi
energy. Experimentally, this is the case: in the two heavy-fermion
metals studied until now, Nernst coefficient is found to be in the
0.2-5$\mu$V/KT range, several orders of magnitude larger than the
isothermal Nernst coefficient of gold($\sim$ 0.1
nV/KT)\cite{bel3}. In the second place, the expression indicates
that any increase in $\Theta_{H}$ (or more fundamentally, in the
quasi-particle lifetime $\tau$) would also lead to an enhancement
of the Nernst coefficient. Interestingly, in both
URu$_{2}$Si$_{2}$ and CeCoIn$_{5}$, the giant Nernst effect
emerges with a concomitant enhancement in the amplitude of
$\Theta_{H}$.
\begin{figure}
\resizebox{!}{0.4\textwidth}{\includegraphics{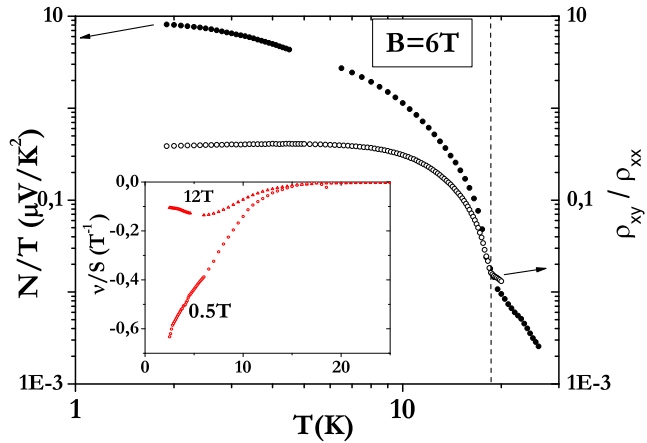}}
\caption{\label{fig4} Nernst coefficient divided by temperature
and the Hall angle as a function of temperature. The dashed
vertical line marks the transition temperature T$_{0}$.Inset: the
temperature dependence of the ratio of the Nernst to Seebeck
coefficients.}
\end{figure}

This brings us to another remarkable feature of electronic
transport in the hidden-order state of URu$_2$Si$_2$. Fig.4
displays the temperature dependence of the Hall angle in a
logarithmic scale. As seen in the figure, the phase transition at
T$_{0}$ leads to an almost thirty-fold increase in the Hall angle.
At B=12T, the system is close to the $\omega_{c}\tau\sim 1$
limit\cite{footnote}. Following the thread of Eq.1, it is
instructive to compare this with the temperature dependence of
$N/T(T)$. As seen in Fig.4, both quantities are considerably
enhanced below T$_{0}$. However, their behavior is not identical.
When the sample is cooled from 17K to 1.5K, $\Theta_H$ increases
by a factor of 30 and saturates rapidly. On the other hand, $N/T$
increases by a factor of 500 and continues to increase smoothly.

Therefore,  a small Fermi Energy combined with a long scattering
time provides a partial explanation for the magnitude of the
Nernst coefficient in URu$_{2}$Si$_{2}$. Numerically, at B=6T and
T=3K, $N \sim 20 \mu V /K$ and $\Theta_H \sim 0.4$ and expression
2 implies $\epsilon_F \sim$ 17K. However, a reduced Fermi energy
would enhance the Nernst and Seebeck coefficients in exactly the
same way. This is, however, not the case. As seen in the inset of
figure, in the low-field regime, upon cooling the ratio $\nu/S$
displays a diverging behavior in a manner analogous to the one
observed in CeCoIn$_{5}$\cite{bel2}. In both cases, the large
Nernst signal is accompanied with an [anomalously] small Seebeck
signal. Moreover, as seen in Fig.4, $N/T$ continues to increase at
low temperature while $\Theta_H$ has already attained a constant
value. These features are yet to be understood. These features
point to a missing ingredient for understanding this enhanced
Nernst signal which may be a generic feature of heavy fermions
physics.

We note that a link between unconventional density wave order and
Nernst signal has been recently proposed\cite{dora}. On the other
hand, given the experimental evidence in favor of electronic phase
separation in the hidden order state of
URu$_{2}$Si$_{2}$\cite{matsuda,chandra2}, one is tempted to
speculate in other directions. For example, magnetic droplets
inside a paramagnetic fluid may couple a magnetic flux line to an
entropy reservoir in an manner similar to superconducting
vortices.

In summary,  our results indicate that the ground state of
URu$_{2}$Si$_{2}$ is one with a low density of itinerant electrons
with large entropy and long lifetime. Most remarkably, we found
that a Nernst signal of exceptionally large magnitude emerges in
this context.


\begin{thebibliography}{99}
\bibitem{footnote} In URu$_{2}$Si$_{2}$, quantum oscillations are easily observable at
magnetic fields well below H$_{c2}$ in crystals with residual
resistivities as large as 10$\mu\Omega$ cm. This is a result of
the unusual combination of the two factors discussed above, namely
the low carrier density and the long quasiparticle lifetime.
\end{thebibliography}
\end{document}